\documentclass[conference,a4paper]{IEEEtran}

\usepackage[utf8]{inputenc} 
\usepackage[T1]{fontenc}
\usepackage{url}
\usepackage{ifthen}
\usepackage{cite}
\usepackage[cmex10]{amsmath} 
\usepackage{amsfonts}
\usepackage{amssymb}
\usepackage{mathtools}
\usepackage{graphicx}
\usepackage{bm}
\usepackage{tikz}

\newtheorem{proposition}{Proposition}
\newtheorem{corollary}{Corollary}

\begin{document}
\title{On the Distribution of Random Geometric Graphs} 
\author{%
  \IEEEauthorblockN{Mihai-Alin~Badiu}
  \IEEEauthorblockA{Department of Electronic Systems\\
  					Aalborg University\\
                    Fredrik Bajers Vej 7, 9220 Aalborg \O st, Denmark\\
                    Email: mib@es.aau.dk}
  \and
  \IEEEauthorblockN{Justin P. Coon}
  \IEEEauthorblockA{Department of Engineering Science\\
                    University of Oxford\\ 
                    Parks Road, Oxford OX1 3PJ, UK\\
                    Email: justin.coon@eng.ox.ac.uk}
}

\maketitle

\begin{abstract}
Random geometric graphs (RGGs) are commonly used to model networked systems that depend on the underlying spatial embedding. We concern ourselves with the probability distribution of an RGG, which is crucial for studying its random topology, properties (e.g., connectedness), or Shannon entropy as a measure of the graph's topological uncertainty (or information content). Moreover, the distribution is also relevant for determining average network performance or designing protocols. However, a major impediment in deducing the graph distribution is that it requires the joint probability distribution of the $n(n-1)/2$ distances between $n$ nodes randomly distributed in a bounded domain. As no such result exists in the literature, we make progress by obtaining the joint distribution of the distances between three nodes confined in a disk in $\mathbb{R}^2$. This enables the calculation of the probability distribution and entropy of a three-node graph. For arbitrary $n$, we derive a series of upper bounds on the graph entropy; in particular, the bound involving the entropy of a three-node graph is tighter than the existing bound which assumes distances are independent. Finally, we provide numerical results on graph connectedness and the tightness of the derived entropy bounds. 
\end{abstract}


\section{Introduction}


Uncertainty is pervasive in modern wireless networks.  The sources of this uncertainty range from the humans that interact with the networks and the locations of the nodes in space down to the transmission protocols and the underlying scattering processes that affect signal propagation.  To date, some progress has been made towards characterizing the structural uncertainty of wireless networks by modeling these networks as random geometric graphs (RGGs) where the probability that two particular nodes are connected is a function of the distance between them~\cite{Franceschetti2008,Baccelli2009,Dettmann2016}.  RGGs with probabilistic pair connection functions are known in the mathematics community as \emph{soft} RGGs~\cite{Penrose2016}.  Work on these graphs has mostly been focused on analyzing their percolation (in an infinite domain) or connectivity (in a finite domain) properties~\cite{Penrose1991,Meester1996,Coon2012}.  In the case of finite (but dense) graphs, this sort of investigation typically amounts to obtaining an understanding of the probability that a single isolated node exists.  

Ideally, one would like to obtain information about the complete distribution of the graphs in the ensemble.  This information would enable us to study not only connectivity, but also important features such as topological structure and complexity through the lens of graph entropy~\cite{Simonyi1995}.  Applications of entropy-based methods to the study of networked systems are abundant and include problems related to molecular structure classification~\cite{Bonchev1983}, social networks~\cite{Everett1985,Dehmer2011}, data compression~\cite{Choi2012}, and quantum entanglement~\cite{deBeaudrap2016,Simmons2017}.  Graph entropy has also been invoked in the study of communication networks to quantify node and route stability~\cite{Guo2011} with the aim of improving link prediction~\cite{Zayani2012} and routing protocols~\cite{An2002,Boushaba2017}.  Topological uncertainty in dynamic mobile ad hoc networks was investigated in~\cite{Timo2005} from a network layer perspective, and~\cite{Lu2008} treated self-organisation in networks using a basic graph entropy framework.  More recently, an analytical approach for studying topological uncertainty in wireless networks was proposed in~\cite{Coon2016,Coon2017,Cika2017}.

In this paper, we study the probability distribution of the RGG formed by $n$ nodes randomly distributed in a bounded domain. The joint distribution of all $n(n-1)/2$ inter-node distances is greatly relevant for the distribution of the RGG. Finding distance distributions is a very challenging task in probabilistic geometry, as it often leads to intractable definite integrals; existing literature focuses on the distance between two nodes or the distances between a node and its neighbours (e.g., see~\cite{Mathai1999,Srinivasa2010,Khalid2013,Pure2015}). We derive the joint distribution of the inter-node distances in closed-form, for $n=3$ nodes confined in a disk in $\mathbb{R}^2$; to our knowledge, this is the first time such a result is obtained. We avoid intractable integrations by using a conditioning technique and expect that the same approach could be used for larger $n$. Also, for arbitrary $n$, we derive a series of upper bounds on the graph entropy; in particular, the bound involving the entropy of a three-node graph is tighter than the existing bound which assumes distances are independent. Finally, we provide numerical results on graph connectedness and the tightness of the derived entropy bounds. 

\section{Random Geometric Graph}
\subsection{Model}
Consider a set $\mathcal{V}_n=\{1,\ldots,n\}$ of $n$ nodes that are randomly located in a space $\mathcal{K}\subset \mathbb{R}^d$ of finite volume and diameter $D\coloneqq \sup_{u,v\in\mathcal{K}}\|u-v\|$. We assume that the locations $\{Z_i\}_{i\in\mathcal{V}_n}$ of the nodes are independently and uniformly distributed in $\mathcal{K}$. The existence of an (undirected) edge between nodes $i$ and $j$ depends on the Euclidean distance between the two nodes and is indicated by the binary random variable $X_{ij}$ being one. Specifically, given the node locations, the variables $\{X_{ij}\}$ are independent and each edge $(i,j)$ exists with probability 
\begin{equation}
P(X_{ij}=1\vert z_i,z_j)=p(\|z_i-z_j\|),
\end{equation}
where $p:[0,\infty)\to[0,1]$ is the pair connection function. For example, in the hard disk model, $p(\cdot)$ is an indicator function that equals one when its argument is less than $r_0$ and zero otherwise, where $r_0$ denotes the maximum connection range. We define the binary vector $\bm{X}_n$ to include all edge variables, i.e., $\bm{X}_n=(X_{ij})_{i<j}$. The random geometric graph $G_n\coloneqq G(\mathcal{V}_n,\mathcal{E}_n)$ with edge set $\mathcal{E}_n=\{(i,j)\mid X_{ij}=1\}$ is distributed in the set of all $n(n-1)/2$ possible graphs. 

\subsection{Probability Distribution and Entropy}
The distribution of $G_n$ is determined by both the distribution of locations $\{Z_i\}_{i\in\mathcal{V}_n}$ and the probabilistic connection model specified by $p(\cdot)$. The graph $G_n$ is uniquely determined by $\bm{X}_n$, which has a multivariate Bernoulli distribution. Therefore, we study the pmf $f_{\bm{X}_n}(\bm{x}_n)\coloneqq P(\bm{X}_n=\bm{x}_n)$, for each $\bm{x}_n\in\{0,1\}^{n(n-1)/2}$. Since the conditional probability of edge existence depends on distance, it is more convenient to work with inter-node distances instead of node locations. Let $\bm{R}_n\coloneqq (R_{ij})_{i<j}$ denote the random vector collecting the pair distances $R_{ij}\coloneqq \|Z_i-Z_j\|$, and let $f_{\bm{R}_n}:[0,D]^{n(n-1)/2}\to[0,\infty)$ be its pdf. We now write 
\begin{multline}\label{eq:pmf_graph}
f_{\bm{X}_n}(\bm{x}_n) \\
=\int_{\mathcal{R}} f_{\bm{R}_n}(\bm{r}_n)  \prod_{\substack{i,j=1 \\ i<j}}^n  p^{x_{ij}}(r_{ij}) \left[1-p(r_{ij})\right]^{1-x_{ij}} \operatorname{d}\!r_{ij}.
\end{multline}
where the integration domain is $\mathcal{R}=[0,D]^{n(n-1)/2}$.
The distribution of $\bm{X}_n$ is symmetric, since the node locations are identically distributed and the pair connection function is the same for all edges. 
The topological uncertainty (or information content) of $G_n$ can be quantified by the Shannon entropy, i.e.,
\begin{align}\label{eq:entropy}
H(G_n)&=H(\bm{X}_n)\nonumber \\
&=-\sum_{\bm{x}_n\in\{0,1\}^{n(n-1)/2}} f_{\bm{X}_n}(\bm{x}_n) \log f_{\bm{X}_n}(\bm{x}_n).
\end{align}

It is clear from~\eqref{eq:pmf_graph} that the joint pdf $f_{\bm{R}_n}$ of inter-node distances is highly important for the graph distribution and its entropy. For $n=2$, the sought pdf reduces to the pdf of the distance between two nodes, which has been extensively studied for various shapes of the embedding space $\mathcal{K}$ (e.g., see~\cite{Mathai1999,Srinivasa2010,Khalid2013,Pure2015}). Obtaining the joint pdf analytically for $n>2$ is very challenging and no such results have been reported previously. In the next section, we make progress by obtaining the joint pdf for $n=3$ in closed-form by using a conditioning technique. This enables the calculation of the pmf~\eqref{eq:pmf_graph} and entropy~\eqref{eq:entropy} for $n=3$, which then can be used to bound the graph entropy when $n>3$, as shown in Sec.~\ref{sec:entropy_bounds}.

\section{Joint PDF of Inter-Node Distances for $n=3$}\label{sec:jointPDF}
We consider $n=3$ and $\mathcal{K}$ is a disk of diameter $D$ in $\mathbb{R}^2$. Even though the locations of the three nodes are independently and uniformly distributed, determining the joint pdf of the three distances by direct integration is very difficult. For example, one could attempt to transform the Cartesian coordinates (i.e., six variables) to other coordinates that include the three distances, apply the transformation theorem and integrate out the redundant coordinates. However, this leads to complicated definite integrals, because triangle inequalities and the condition that the points have to be inside the circle need to be ensured. 

Computing integrals over complicated regions is often required in probabilistic geometry. Crofton's technique~\cite{Solomon1978} has proven to simplify such evaluations in many problems, such as finding the distribution of the distance between two random points~\cite{Alagar1976}. The work~\cite{Eisenberg2000} shows that Crofton's method is essentially equivalent to the technique of computing expectations by conditioning. We use the latter in the following. 

Our approach is to compute the joint pdf conditioned on an additional (suitably chosen) random variable, which is easier than the original problem. Then, we obtain the desired joint pdf by taking the expectation of the conditional pdf over the density of the additional variable. We expect that this approach is also useful for $n>3$. 

Before presenting the result, we fix some notation. For a triangle with side lengths $r_{12}$, $r_{13}$ and $r_{23}$, let $d$ be the diameter of its circumscribed circle, i.e.,
\begin{equation}\label{eq:circumdiameter}
d = \frac{2r_{12}r_{13}r_{23}}{\sqrt{Q(r_{12},r_{13},r_{23})}},
\end{equation}
where $Q(r_{12},r_{13},r_{23})=2r_{12}^2r_{13}^2+2r_{12}^2r_{23}^2+2r_{13}^2r_{23}^2 - r_{12}^4-r_{13}^4-r_{23}^4$; note that $Q(r_{12},r_{13},r_{23})>0$ is equivalent to $r_{12}$, $r_{13}$, $r_{23}$ satisfying the triangle inequalities. We denote the largest side length by $\bar{r}=\max(r_{12},r_{13},r_{23})$. Let us also define the function $\varphi:[0,1]\to \mathbb{R}$, $\varphi(x)=\arccos(x) -x\sqrt{1-x^2}$. 

\begin{proposition}\label{prop:pdf}
Assume three points are independently and uniformly distributed inside a circle of diameter $D$ and let $R_{12}$, $R_{13}$ and $R_{23}$ be the side lengths of the random triangle determined by the points. Then, for all $r_{12}$, $r_{13}$, $r_{23}\in\mathbb{R}_{+}$ such that $Q(r_{12},r_{13},r_{23})>0$ and $\bar{r}\leq D$, the joint pdf of the side lengths  is given by eq.~\eqref{eq:jointPDF} at the top of the next page. The pdf depends on whether the realized triangle is obtuse or acute, and whether the diameter~\eqref{eq:circumdiameter} of its circumscribed circle is larger or smaller than $D$. 
\end{proposition}
\begin{IEEEproof}
An outline of the proof is given in the appendix.
\end{IEEEproof}
\begin{figure*}
\begin{equation}\label{eq:jointPDF}
f_{\bm{R}_3}(r_{12},r_{13},r_{23})
=
\begin{cases}
\frac{64d}{\pi^2 D^4}\left\{\sum_{i<j} \left[ \varphi\left(\frac{r_{ij}}{D}\right) - \frac{d^2}{D^2}\varphi\left(\frac{r_{ij}}{d}\right) \right] - \frac{\pi}{2}\left(1-\frac{d^2}{D^2}\right) + 2\frac{d^2}{D^2}\varphi\left(\frac{\bar{r}}{d}\right)\right\}, &\text{if } 2\bar{r}^2>\sum_{i<j} r_{ij}^2, \quad d\leq D,\\
\frac{64d}{\pi^2 D^4}\left\{\sum_{i<j} \left[ \varphi\left(\frac{r_{ij}}{D}\right) - \frac{d^2}{D^2}\varphi\left(\frac{r_{ij}}{d}\right) \right] - \frac{\pi}{2}\left(1-\frac{d^2}{D^2}\right) \right\}, &\text{if } 2\bar{r}^2\leq \sum_{i<j} r_{ij}^2, \quad d\leq D,\\
\frac{128d}{\pi^2 D^4} \varphi\left(\frac{\bar{r}}{D}\right), &\text{if } 2\bar{r}^2>\sum_{i<j} r_{ij}^2,\quad d>D, \\
0, &\text{if } 2\bar{r}^2\leq \sum_{i<j} r_{ij}^2, \quad d>D.\\
\end{cases}
\end{equation}
\end{figure*}

\section{Bounding the Graph Entropy}\label{sec:entropy_bounds}
In~\cite{Coon2016,Coon2017}, the upper-bound $H(G_n)\leq{n\choose 2}H(G_2)$ is obtained for any $n\geq 2$ by assuming that $\{X_{ij}\}$ are independent (or, equivalently, that the pair distances $\{R_{ij}\}$ are independent). While such an upper bound is simple and amenable to further analysis, its tightness might not always be sufficient. We set out to find tighter upper bounds by trying to preserve the dependency between pair distances. First, we establish the following result. 
\begin{proposition} For any $m,n\in\mathbb{Z}$ such that $n>m\geq 2$, the entropies of $G_n$ and $G_m$ are related by
\begin{equation}\label{eq:entropy_ineq}
\frac{H(G_n)}{n(n-1)} \leq \frac{H(G_m)}{m(m-1)}.
\end{equation} 
\end{proposition}
\begin{IEEEproof}
The entropy of $G_n$ is given by the entropy of the ${n\choose 2}$ binary variables in $\bm{X}_n$, see~\eqref{eq:entropy}. Our intention is to relate $H(G_n)$ to the entropy of RGGs with smaller number of nodes. Specifically, for $m<n$, we consider all the ${n\choose m}$ subsets of $\mathcal{V}_n$ that have $m$ nodes. Let $\mathcal{N}_{m,k}\subset\mathcal{V}_n$ be the $k$th such subset, $k=1,\ldots,{n\choose m}$. The set of pair indices corresponding to $\mathcal{N}_{m,k}$ is denoted by $S_k=\{ij\mid i,j\in\mathcal{N}_{m,k}, i<j \}$. We further define the set $\mathcal{S}=\{S_1,\ldots,S_{{n\choose m}}\}$ collecting all the sets of pair indices. In this construction, each pair index $ij$ with $i,j\in\mathcal{V}_n$ appears in ${{n-2}\choose{m-2}}$ subsets of $\mathcal{S}$. According to Shearer's inequality, which is a generalization of the subadditivity of joint entropy~\cite{Chung1986,Yanagida2008}, we have
\begin{equation}\label{eq:Shearer_ineq}
	H(\bm{X}_n) \leq \frac{1}{{{n-2}\choose{m-2}}}\sum_{S\in\mathcal{S}} H(\bm{X}_S),
\end{equation}
where $\bm{X}_S\coloneqq(X_{ij})_{ij\in S}$. Each term in the r.h.s. of~\eqref{eq:Shearer_ineq} is the entropy of a graph with $m$ nodes; by invoking the system's symmetry, all terms are equal to $H(G_m)$, such that
\begin{equation}
	H(G_n)\leq  \frac{{{n}\choose{m}}}{{{n-2}\choose{m-2}}} H(G_m),
\end{equation}
and~\eqref{eq:entropy_ineq} follows immediately.  
\end{IEEEproof}
The following corollary gives a series of tighter and tighter upper bounds on $H(G_n)$, for all $n\geq 2$.
\begin{corollary} The normalized (i.e., per edge) entropy decreases with the number of nodes, i.e.,
\begin{equation}\label{eq:entropy_bounds}
	\frac{H(G_n)}{{{n}\choose{2}}} \leq  \frac{H(G_{n-1})}{{{n-1}\choose{2}}} \leq \ldots \leq \frac{H(G_3)}{3} \leq H(G_2).
\end{equation}
\end{corollary}
\begin{IEEEproof}
We immediately obtain~\eqref{eq:entropy_bounds} by successively applying~\eqref{eq:entropy_ineq} for consecutive integers.
\end{IEEEproof}

\section{Numerical Experiments}
In the following we assume that the random nodes are confined in a disk with diameter $D=1$; any two nodes are connected by an edge if and only if the distance between them is less than $r_0$. 

\begin{figure}[t]
   \centering
   \includegraphics[width=\columnwidth]{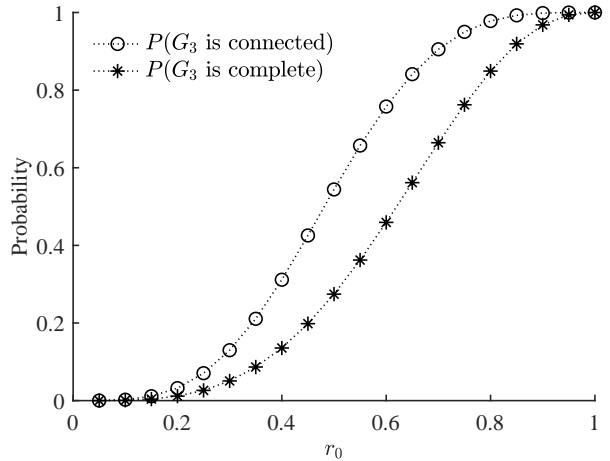}
   \caption{Probability of connectedness and probability of completeness for an RGG with $n=3$ nodes and maximum connection range $r_0$; the three nodes are randomly located inside a circle with diameter one.}
   \label{fig:Prob}
\end{figure}

We first take an example from ad-hoc communications, where it is relevant to know conditions under which any two nodes of the network can communicate. If multi-hop communication is possible, this is equivalent to the requirement that the graph be connected; otherwise, the graph needs to be complete. We consider a three-node graph and evaluate $P(G_3\text{ is connected})=f_{\bm{X}_3}(0,1,1)+f_{\bm{X}_3}(1,0,1)+f_{\bm{X}_3}(1,1,0)+f_{\bm{X}_3}(1,1,1)$ and $P(G_3\text{ is complete})=f_{\bm{X}_3}(1,1,1)$ as functions of $r_0$ (which can be thought of as being monotonically related to the transmit power). We compute the pmf~\eqref{eq:pmf_graph} by using the derived joint pdf~\eqref{eq:jointPDF} and numerical integration. The results in Fig.~\ref{fig:Prob} show that two-hop relaying significantly improves the probability that any two of the three nodes can communicate.  

\begin{figure}[t]
   \centering
   \includegraphics[width=\columnwidth]{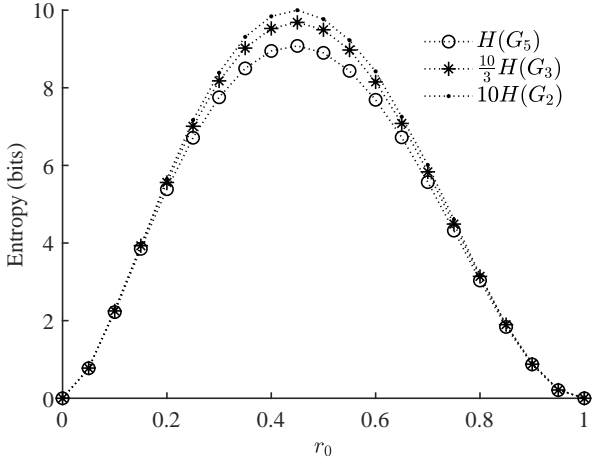}
   \caption{Entropy of an RGG with $n=5$ nodes and maximum connection range $r_0$, and upper bounds; the five nodes are randomly located inside a circle with diameter one.}
   \label{fig:Entropy}
\end{figure}

We now study the entropy bounds derived in Sec.~\ref{sec:entropy_bounds}. We consider $n=5$ nodes and compute $H(G_5)$ using Monte Carlo simulation. From~\eqref{eq:entropy_bounds}, we have $H(G_5)\leq\frac{10}{3}H(G_3)\leq 10H(G_2)$. We use the derived joint pdf~\eqref{eq:jointPDF} to compute the pmf~\eqref{eq:pmf_graph}, which then gives $H(G_3)$. We similarly obtain $H(G_2)$ based on the pdf of the distance between two points inside a circle~\cite{Mathai1999}. Fig.~\ref{fig:Entropy} shows that $H(G_5)$ approaches zero when $r_0\to 0$ or $r_0\to D$ (i.e., when the RGG becomes deterministically empty or complete, respectively). The entropy is significant at intermediate values of $r_0$ and always less than $10$ bits, which is the entropy of a five-node graph whose $10$ potential edges exist independently with probability $0.5$. We can also observe that the bound based on $H(G_3)$ provides an improvement over the one obtained by assuming the $10$ inter-node distances are independent. 

\section{Conclusion}
In this paper, we studied the distribution of a random geometric graph and its entropy. The distribution provides insights into properties of the random graph, such as topological structure or connectivity, while entropy is useful for understanding topological complexity. We showed that the normalized (per edge) entropy decreases with the number of nodes. This result gave a series of upper bounds on entropy, each bound involving the entropy of a graph with smaller number of nodes. We pointed out the importance of the joint distribution of pair distances in determining the graph's distribution and its entropy, and the lack of such results in the literature. We progressed by deriving the joint distribution of distances between three nodes confined in a disk, and expect that the approach we used could be applied for larger number of nodes. 

\appendix
Let $O$ be the center of the disk $\mathcal{K}$ of diameter $D$. We denote by $S_i$ the minimum diameter of a disk centred at $O$ that includes the $i$th point and define $\bar{S}=\max(S_1,S_2,S_3)$. We write
\begin{equation}\label{eq:cond2jointPDF}
f_{\bm{R}_3}(r_{12},r_{13},r_{23})=\int_0^D f_{\bm{R}_3\mid \bar{S}}(r_{12},r_{13},r_{23}\mid s) f_{\bar{S}}(s)\operatorname{d}\!s.
\end{equation}
Conditioning on $\bar{S}$ is very convenient because, in the computation of $f_{\bm{R}_3\mid \bar{S}}$, one of the three points is on the circle $\mathcal{C}_s$ of center $O$ and diameter $s$, while the other two points are inside $\mathcal{C}_s$; this is a great simplification. The density $f_{\bar{S}}$ is obtained as follows: we have $P(S_i\leq s)=s^2/D^2$, for each $s\in[0,D]$; therefore, $P(\bar{S}\leq s)=s^6/D^6$, which gives the pdf $f_{\bar{S}}(s)=6s^5/D^6$. 

To compute $f_{\bm{R}_3\mid \bar{S}}$, we study the ``number of ways'' in which one can fit a triangle of side-lengths $r_{12}$, $r_{13}$ and $r_{23}$ inside $\mathcal{C}_s$ when one of the triangle's vertices is fixed on the circle. The side lengths must satisfy the triangle inequalities, which is equivalent to $Q(r_{12},r_{13},r_{23})\coloneqq 2r_{12}^2r_{13}^2+2r_{12}^2r_{23}^2+2r_{13}^2r_{23}^2 - r_{12}^4-r_{13}^4-r_{23}^4>0$. It is also required that $\bar{r}\coloneqq\max(r_{12},r_{13},r_{23})\leq s$. 

\begin{figure}[t]
\centering
\begin{tikzpicture}[scale=1.66]
\draw(0,0) circle [radius=2cm];
\coordinate (o) at (0,0);
\coordinate (a) at (180:2cm);
\coordinate (b) at (60:1.2cm);
\coordinate (c) at (275:1cm);
\node(O)[label=$O$] at (o) {.};
\node(A)[label=left:$A_i$] at (a) {$\bullet$};
\node(B)[label=right:$A_j$] at (b) {$\bullet$};
\node(C)[label=below:$A_k$] at (c) {$\bullet$};
\draw[dotted] (a) -- (o);
\draw (a) -- (b) -- (c) -- cycle;
\draw[dotted] ([shift=(314.427:2.8cm)]180:2cm)  arc (-45.573:45.573:2.8cm);
\draw[dotted] (a) -- ([shift=(45.573:2.8cm)]180:2cm);
\draw[->] (a)+(-25:0.3cm) arc (-25:23:0.3cm);
\node[] at ([shift=(-3:0.5cm)]180:2cm) {$\Theta_{i}$};
\draw[->] (a)+(0:0.75cm) arc (0:22.5:0.75cm);
\node[] at ([shift=(9:0.95cm)]180:2cm) {$\Theta_{ij}$};
\draw[->] (a)+(0:1.2cm) arc (0:45.573:1.2cm);
\node[] at ([shift=(32:1.4cm)]180:2cm) {$\bar{\theta}_{ij}$};
\end{tikzpicture}
\caption{Illustration of the circle $\mathcal{C}_s$ of center $O$ and diameter $s$; the point $A_i$ is on the circle, while $A_j$ and $A_k$ are inside $\mathcal{C}_s$.}
\label{fig:circ}
\end{figure}
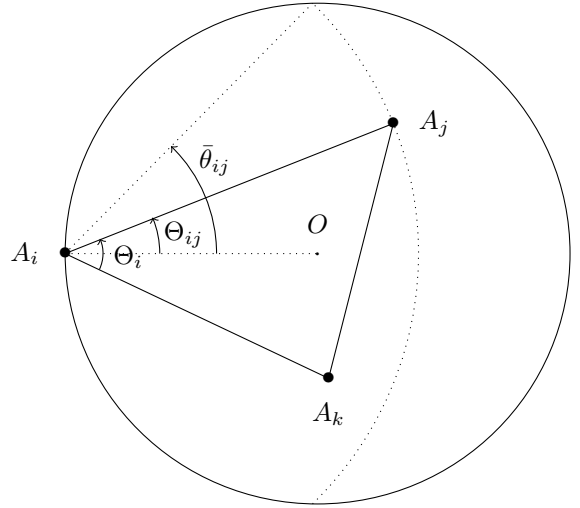

In Fig.~\ref{fig:circ}, point $A_i$ represents node $i$. Assuming $A_i$ is on $\mathcal{C}_s$ we have
\begin{align}\label{eq:jointPDF_i_on_circ}
&f_{\bm{R}_3}^i(r_{12},r_{13},r_{23}) \nonumber\\
&= f_{R_{jk}\vert R_{ij},R_{ik}}^i(r_{jk}\vert r_{ij},r_{ik}) f_{R_{ij}}^i(r_{ij})f_{R_{ik}}^i(r_{ik})
\end{align}
where superscript $i$ indicates conditioning on node $i$ being on $\mathcal{C}_s$, and $\{i,j,k\}\equiv\{1,2,3\}$. For each $j\neq i$, the pdf of $R_{ij}=|A_iA_j|$ is~\cite{Mathai1999}
\begin{equation}\label{eq:pdf_dist_i_on_circ}
f_{R_{ij}}^i(r_{ij}) = \frac{8r_{ij}}{\pi s^2}\arccos\left(\frac{r_{ij}}{s}\right),\quad r_{ij}\in[0,s].
\end{equation}
To obtain $f_{R_{jk}\vert R_{ij},R_{ik}}^i$, we use the law of cosines $R_{jk}^2 = R_{ij}^2+R_{ik}^2 -2R_{ij}R_{ik}\cos\Theta_i$, with $\Theta_i\coloneqq \angle A_jA_iA_k$. For each $j\neq i$, we further define $\Theta_{ij}=\angle OA_iA_j$; we have $\Theta_{ij}\vert R_{ij}\sim \mathcal{U}(-\bar{\theta}_{ij},\bar{\theta}_{ij}))$, with $\bar{\theta}_{ij}=\arccos\left(\frac{r_{ij}}{s}\right)<\pi/2$. Since $\Theta_i = \Theta_{ij} - \Theta_{ik}$ (i.e., the difference between two independent and uniformly distributed variables), it follows that $\Theta_i\vert R_{ij},R_{ik}$ has a trapezoidal distribution with pdf
\begin{align*}
&f_{\Theta_i\vert R_{ij},R_{ik}}^i(\theta_i\vert r_{ij},r_{ik}) \\
&= \begin{cases}
\frac{1}{2\max(\bar{\theta}_{ij},\bar{\theta}_{ik})}, &\text{if } 0\leq |\theta_i|\leq |\bar{\theta}_{ij}-\bar{\theta}_{ik}|,\\
\frac{\bar{\theta}_{ij}+\bar{\theta}_{ik}-|\theta_i|}{4\bar{\theta}_{ij}\bar{\theta}_{ik}}, &\text{if } |\bar{\theta}_{ij}-\bar{\theta}_{ik}| \leq |\theta_i|< \bar{\theta}_{ij}+\bar{\theta}_{ik}, \\
0, &\text{if } \bar{\theta}_{ij}+\bar{\theta}_{ik} \leq |\theta_i| <\pi.
\end{cases}
\end{align*}
Now, we make the transformation $Y=\cos\Theta_i$ and obtain the pdf of $Y$ from its cdf, which is computed as $F_{Y\vert R_{ij},R_{ik}}^i(y\vert r_{ij},r_{ik})=1-P(\cos\Theta_i>y)$. Then, we use the law of cosines and transformation theorem to obtain $f_{R_{jk}\vert R_{ij},R_{ik}}^i$. We distinguish between several cases depending on the diameter $d$ of the circumscribed circle~\eqref{eq:circumdiameter}. When $d>s$, the only way in which the triangle can be inside $\mathcal{C}_s$ while node $i$ is on $\mathcal{C}_s$ is when the triangle is obtuse (i.e., $2\bar{r}^2> r_{12}^2+r_{13}^2+r_{23}^2$) and its largest side length $\bar{r}$ is either $r_{ij}$ or $r_{ik}$. Using~\eqref{eq:pdf_dist_i_on_circ} in~\eqref{eq:jointPDF_i_on_circ}, we obtain
\begin{align}\label{eq:jointPDF_i_circ_form}
&f_{\bm{R}_3}^i(r_{12},r_{13},r_{23}) \nonumber\\
&=
\begin{cases}
\frac{32d}{\pi^2s^4}(\bar{\theta}_{ij}+\bar{\theta}_{ik}-\theta_i), & \text{if } d\leq s, \\
\frac{64d}{\pi^2s^4} \arccos\left(\frac{\bar{r}}{s}\right), & \text{if } d>s, \quad 2\bar{r}^2>\sum_{i<j}r_{ij}^2, \\
0, & \text{else},
\end{cases} 
\end{align}
for all $r_{12}$, $r_{13}$, $r_{23}\in\mathbb{R}_{+}$ that satisfy $Q(r_{12},r_{13},r_{23})>0$ and $\bar{r}\leq s$. Since each node can be on the circle with probability $1/3$, it follows that $f_{\bm{R}_3\mid \bar{S}}=\sum_{i=1}^3 f_{\bm{R}_3}^i/3$, which gives
\begin{align}\label{eq:condPDF}
&f_{\bm{R}_3\mid \bar{S}}(r_{12},r_{13},r_{23}\vert s) \nonumber\\
&=
\begin{cases}
\frac{64d}{3\pi^2s^4}\left[\sum\limits_{i<j}\arccos\left(\frac{r_{ij}}{s}\right) - \frac{\pi}{2}\right], & \text{if } d\leq s, \\
\frac{128d}{3\pi^2s^4}  \arccos\left(\frac{\bar{r}}{s}\right), & \text{if } d>s,  2\bar{r}^2>\sum\limits_{i<j}r_{ij}^2, \\
0, & \text{else}.
\end{cases} 
\end{align}
We have used that $\theta_1+\theta_2+\theta_3=\pi$ and when $d>s$ the node corresponding to the obtuse angle cannot be on $\mathcal{C}_s$. 

Finally, by plugging~\eqref{eq:condPDF} into~\eqref{eq:cond2jointPDF}, we calculate the integral by distinguishing between the cases $d\leq D$ and $d>D$, and arrive at the closed-form expression~\eqref{eq:jointPDF}.

\section*{Acknowledgment}
This work was supported by Independent Research Fund Denmark grant number DFF–5054-00212 and by EPSRC grant number EP/N002350/1 (``Spatially Embedded Networks''). The research was carried out during a visit to University of Oxford, and M.-A Badiu would like to thank the Communications Research Group for the hospitality.


\bibliographystyle{IEEEtran}
\bibliography{refs}

\end{document}